# Combinatorial Reactive Sputtering with Auger Parameter Analysis enables Synthesis of Wurtzite $Zn_2TaN_3$.

Siarhei Zhuk[a], Alexander Wieczorek[a], Amit Sharma[b], Jyotish Patidar,[a]

Kerstin Thorwarth[a], Johann Michler[b], Sebastian Siol[a,*]

a) Empa – Swiss Federal Laboratories for Materials Science and Technology, 8600 Dübendorf, Switzerland

b) Empa – Swiss Federal Laboratories for Materials Science and Technology, 3602 Thun, Switzerland

*E-mail: sebastian.siol@empa.ch

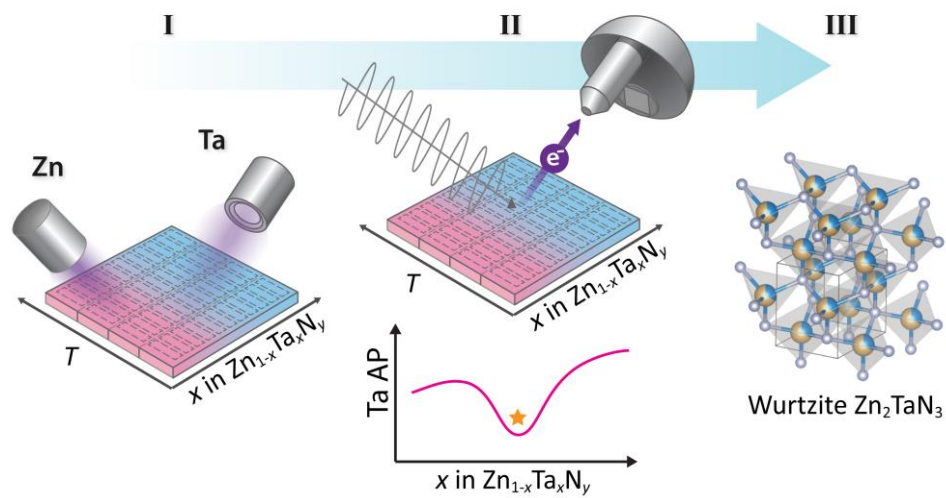

## Abstract

The discovery of new functional materials is one of the key challenges in materials science. Combinatorial high-throughput approaches using reactive sputtering are commonly employed to screen unexplored phase spaces. During reactive combinatorial deposition the process conditions are rarely optimized, which can lead to poor crystallinity of the thin films. In addition, sputtering at shallow deposition angles can lead to off-axis preferential orientation of the grains. This can make the results from a conventional structural phase screening ambiguous. Here we perform a combinatorial screening of the Zn-Ta-N phase space with the aim to synthesize the novel semiconductor $Zn_2TaN_3$. While the results of the XRD phase screening are inconclusive, including Auger parameter analysis in our workflow allows us to see a very clear discontinuity in the evolution of the Ta binding environment. This is indicative of the formation of a new ternary phase. In additional experiments, we isolate the material and perform a detailed characterization confirming the formation of single-phase wurtzite $Zn_2TaN_3$. Besides the formation of the new ternary nitride, we map the functional properties of $Zn_xTa_{1-x}N$ and report previously unreported clean chemical state analysis for $Zn_3N_2$, TaN and $Zn_2TaN_3$. Overall, the results of this study showcase common challenges in high-throughput materials screening and highlight the merit of employing characterization techniques sensitive towards changes in the materials' short-range order and chemical state.

**Introduction**

The discovery and development of new functional materials is a key challenge in the development of next-generation technologies. Nitrides are an important class of functional materials with considerable innovation potential for many applications.[1–3] One prominent example is In$_x$Ga$_{1-x}$N, which played an important part in igniting the solid-state lighting revolution.[4,5] Despite their undisputable technological relevance, nitrides remain a rather underexplored class of materials, especially when compared to their oxide and sulfide counterparts.[6] Recent works on the high-throughput computational prediction of novel stable and metastable ternary metal nitrides clearly show that many potentially exciting functional nitrides are yet to be experimentally realized.[7] One potential reason for this is the challenging synthesis of phase-pure nitrides as they are often metastable with respect to their precursors and prone to oxygen contamination.[8]

Fueled by advances in high-throughput computation as well as experimental infrastructure, the rate of discovery for novel nitrides has increased substantially in the last decade. Non-equilibrium physical vapor deposition (PVD) in ultra-high vacuum (UHV) conditions provides ideal synthesis conditions for metastable nitrides. At the same time, combinatorial synthesis, automated characterization and advanced data analysis techniques are being used to accelerate the screening of complex phase spaces.[9–12] By rapidly covering large windows of the synthesis parameter space, new phases with narrow stability windows can easily be identified and isolated. High-throughput computational predictions combined with combinatorial PVD screening experiments already resulted in the discovery of numerous novel ternary nitride compounds.[3,6,7,13,14].

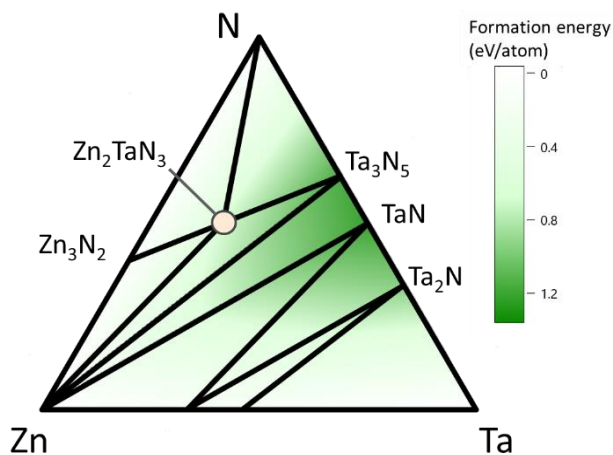

**Fig. 1.** Simulated ternary phase diagram of the Zn-Ta-N material system at 0 K. Zn$_2$TaN$_3$ is predicted to be thermodynamically stable with a low stability w.r.t. to decomposition. The formation enthalpy of TaN is low compared to other compounds in the phase space.[15]

A class of materials within this space, which is currently receiving considerable attention, are ternary zinc nitrides with the formula Zn$_2$MN$_3$. We recently reported on the synthesis of Zn$_2$VN$_3$, whereas several other new ternary zinc nitrides have been successfully synthesized including Zn$_2$SbN$_3$ and Zn$_2$NbN$_3$.[14,16–18] Another promising candidate among the ternary zinc nitrides, which has not been experimentally demonstrated yet, is Zn$_2$TaN$_3$.[6,15,19] In a recent comprehensive review on ternary metal nitrides, Zakutayev *et al.* reported that cation-ordered Zn$_2$TaN$_3$ is predicted to crystallize in an orthorhombic crystal structure (SG36). This phase has a low formation enthalpy of $\Delta H_f$ = –0.63 eV formation enthalpies with a thermochemical stability w.r.t. decomposition of $\Delta H_D$ = –0.16 eV. In addition, a low-energy rocksalt polymorph is predicted at 0.167 eV/atom above the ground state.[6] It is important to note, that several ternary zinc nitrides have been predicted to crystallize in a cation-ordered orthorhombic structure (SG36), whereas the synthesized thin films often exhibit pronounced cation-disorder rendering the structure more similar to hexagonal wurtzite (SG186).[14,20,21]

The ground state orthorhombic Zn$_2$TaN$_3$ phase has a calculated direct band gap of 3.87 eV and low effective masses for electrons and holes of 1.32 and 3.66, respectively.[6] Hinuma *et al.* predicted similar values of 3.58 eV using Heyd–Scuseria–Ernzerhof (HSE) simulations.[19] These properties make it an interesting, non-toxic, earth-abundant candidate for various electronic applications. Despite these comprehensive computational efforts no other competing ternary phases are predicted in the Zn-Ta-N phase space as shown in Fig. 1, which should facilitate the synthesis of single-phase $Zn_2TaN_3$.[15,19] Strikingly, despite its predicted stability and a considerable interest in the community, no experimental demonstration of single-phase $Zn_2TaN_3$ has been reported to date.

The reason for this might not be, that the material itself is intrinsically unstable, but rather that its synthesis is particularly difficult. Even though modern high-throughput experimental workflows can significantly accelerate the discovery of new materials, probing the synthesizability of predicted compounds remains difficult.[22,23] In particular, it might be difficult to rule out false negatives during experimental phase screening. In many groups, the typical approach for combinatorial (experimental) phase screening of new thin film materials, is primarily based on automated X-ray diffraction (XRD) and X-ray Fluorescence (XRF) characterization tools.[12,24–30] However, what if the crystallinity of the synthesized thin films is low or the results are ambiguous? Potentially, this might lead to the discarding of prospective material systems from further studies if the combined XRD/XRF analysis does not show clear evidence of a novel phase forming. This is particularly relevant for data-driven materials design approaches, where low crystallinity poses significant challenges for unassisted phase identification.[31] Yet, the problem of low crystallinity may be due to many reasons including the potential decomposition of a metastable material or poor crystallinity caused by a lack of optimization of the deposition parameters. In the present case, crystallization of Zn-M-N phases is particularly challenging due to the high vapor pressure of Zn, which limits the PVD growth to low synthesis temperatures. [32] The low synthesis temperatures prohibit the formation of larger nuclei, which might necessitate a second annealing step. Zakutayev *et al.* recently demonstrated that AlN capping layers can be used to facilitate post-deposition annealing while preventing oxidation or Zn-loss.[33] Secondly, the films might exhibit low crystallinity due to cation/anion off-stoichiometry.[34] This is especially challenging since common high-throughput methods for composition analysis, such as conventional energy dispersive (ED) XRF, have limitations regarding the quantification of nitrogen owing to its low atomic mass. Thirdly, high-throughput XRD screening of samples with preferential out-of-plane ori-entation can complicate phase identification when grains are aligned out of the diffraction plane. In those cases certain reflections might not show up in the diffraction pattern, leading to apparent differences between the measured and predicted XRD patterns. This is often the case in confocal sputter geometries, which are common in combinatorial gradient deposition.[35,36] However, wide-angle X-ray scattering experiments (*e.g.* 2D-XRD or grazing incidence wide angle X-ray scattering (GIWAXS)) can to some degree alleviate this issue, by covering a wider range in χ-space and therefore capturing additional reflection peaks. Overall, phase identification in novel material systems using standard high-throughput characterization methodology can be challenging for thin films with pronounced texture or low crystallinity. Therefore, complementary characterization techniques suitable for the analysis of samples with low crystallinity can facilitate a comprehensive phase-screening of combinatorial thin film libraries. We recently reported that chemical state analysis based on the Auger parameter (AP) concept can be a powerful complementary tool for the combinatorial screening of new thin film materials.[37] The results indicate that the technique could be particularly useful for the investigation of underexplored material systems with low crystallinity.

In this work, we use a high-throughput combinatorial screening workflow for the experimental exploration of the Zn-Ta-N phase space. Specifically, we use a combination of combinatorial UHV-transfer chemical state analysis in combination with conventional XRF and XRD screening to probe the phase formation during reactive co-sputtering of Zn-Ta-N. Whereas initial XRD studies are inconclusive, the chemical state analysis mapping indicates an abrupt change in the Ta chemical state around the compositional range of $Zn_2TaN_3$. Based on the insights from the AP study, more efforts were undertaken to improve the crystallinity of the material leading to the first experimental demonstration of single-phase wurtzite $Zn_2TaN_3$. Microstructural analysis confirms the structure and composition of the new phase but also reveals a very narrow stability region. Overall these results highlight the importance of including characterization techniques, which are sensitive to short-range order in the materials-screening workflow.

**Experimental**

1.1 mm thick borosilicate glass (EXG) substrates of 50.8 mm × 50.8 mm size were ultrasonically cleaned in acetone, and ethanol, followed by drying in a flow of $N_2$. Then, the substrates were loaded through the load lock into the main chamber of an AJA 1500-F sputtering system with a base pressure of $1 \cdot 10^{-6}$ Pa. $Zn_xTa_{1-x}N$ thin films were prepared by reactive radio frequency (RF) co-sputtering of Zn and Ta targets (50.8 mm in diameter, 99.99% purity each) from two opposing guns. The reactive gas was supplied directly to the chimneys of the sputter guns to facilitate target poisoning. The deposition was carried out at a pressure of 0.8 Pa in a mixed atmosphere of Ar (19.5 sccm) and $N_2$ (10.5 sccm). The sputtering power was varied in the range from 24 W to 40 W and from 38 W to 57 W for Zn and Ta guns, respectively.

Combinatorial $Zn_xTa_{1-x}N$ sample libraries were fabricated to accelerate the screening of Zn-Ta-N phase space. The sputter deposition was carried out in confocal sputter-up geometry with oblique deposition angles onto a static substrate to obtain compositional gradients across the sample library. In addition, a custom-designed substrate holder was used to apply a temperature gradient in an orthogonal direction relative to the composition gradient. The temperature gradient over the sample library is achieved as only 20% of the substrate is in direct contact with the heated holder. This results in a nominal temperature gradient from approximately 150 °C to 70 °C degrees on the hot and cold sides of the substrate, respectively. To minimize additional heating of the substrate, the plasma was confined between two opposing guns using a confocal magnetic configuration of the unbalanced magnetrons. Sputter deposition of phase pure $Zn_2TaN_3$ thin films was performed on rotating substrates.

A Bruker D8 X-ray diffraction (XRD) system with Cu Kα X-ray radiation and Bragg-Brentano geometry was employed for the mapping of structural properties of $Zn_xTa_{1-x}N$ thin films. A Fischer XDV-SDD X-ray fluorescence (XRF) system equipped with a Rh X-ray source was used to determine the Zn to Ta ratio and thickness of the $Zn_xTa_{1-x}N$ thin films. Furthermore, the measured composition was calibrated for the data obtained using 13 MeV 127-I Elastic Recoil Detection analysis (ERDA). A PHI Quantera X-ray photoelectron spectroscopy (XPS) system equipped with monochromatic Al-Kα X-ray radiation was used for the analysis of the surface composition and chemical state of the thin films. The samples were transferred from the sputtering chamber in a custom-built mobile transfer device without breaking vacuum ($p < 5\cdot10^{-7}$ mbar). Charge compensation was performed using electron and ion beam neutralizers. The photoelectron detection angle was maintained at 45°. The main component of adventitious carbon was used for referencing of binding energy scale at 284.8 eV with a typical error of ±0.2 eV in a given material system.[38] It is important to note that this error does not affect the reported Auger parameter (AP) values.[39] The focus was given to the AP of the constituent elements; therefore, respective photoemission and Auger electron spectra were recorded. The cross-section TEM samples were prepared using Ga$^+$ in a Tescan (Lyra3) dual-beam FIB-SEM microscope. Cross-sectional transmission electron microscopy analysis of $Zn_2TaN_3$ thin films was carried out in a Themis 200 G3 spherical aberration (probe) corrected TEM using an accelerating voltage of 200 keV equipped with high-angle annular dark field (HAADF) detector, energy dispersive spectrometer (EDS), and selected area electron diffraction (SAED).

Transmittance and reflectance spectra were recorded using a home-built automated UV-Vis-NIR mapping spectrophotometer that employs deuterium UV and tungsten-halogen broad-band light sources as well as CCD spectrometers from Ocean Insight. Details on the calculation of the absorption coefficient are given in supplementary information. The electrical properties of $Zn_xTa_{1-x}N$ thin films were studied using a custom-designed automated four-point probe measurement system. The system is capable of measuring resistivities in the GΩ-range. For single-phase samples, two point measurements with different setups were attempted to estimate the resistivity.

### Results and discussion

### Phase screening

Thin films with compositions ranging from $Zn_{0.15}Ta_{0.85}N$ to $Zn_3N_2$ were synthesized by reactive RF co-sputtering for the high-throughput combinatorial study. To illustrate the phase evolution in the Zn-Ta-N material system, combined XRD and XRF data are presented in the form of false color plots for samples synthesized at 150 °C and <83 °C as can be seen in Fig. 2a and Fig. 2b, respectively.

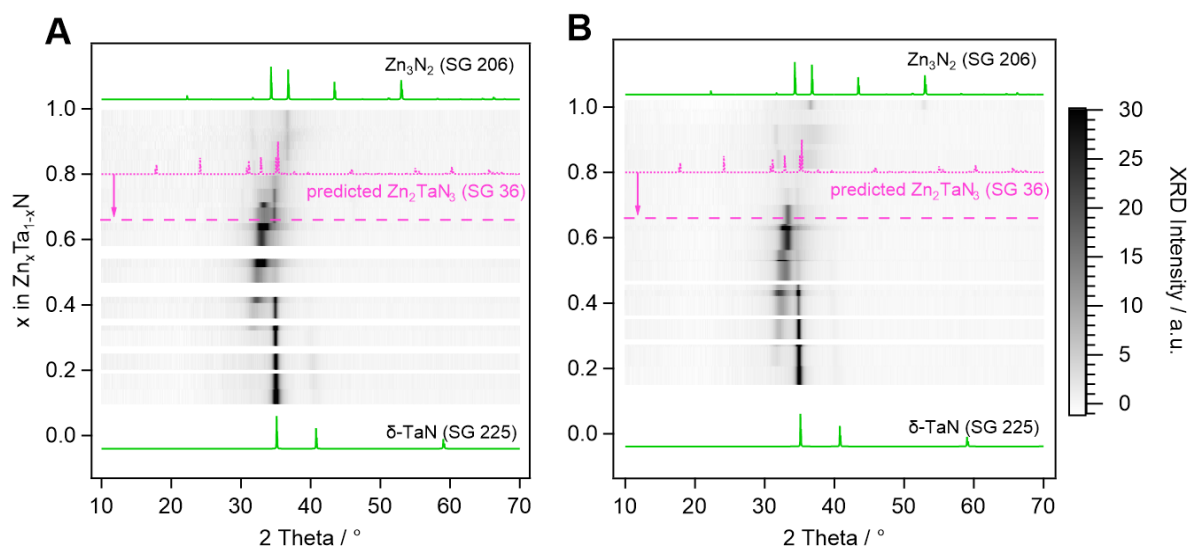

**Figure 2:** XRD and XRF screening of Zn$_x$Ta$_{1-x}$N thin films synthesized at substrate temperatures of (a) 150 °C and (b) <83 °C. At intermediate alloying concentration an additional reflection is observed that indicates the potential formation of a new phase. The low crystallinity of the film, as well as the proximity to signals from competing phases, complicate the analysis.

The plots demonstrate the formation of expected Zn$_3$N$_2$ and TaN phases at opposite sides of the phase diagram. They are in good agreement with reference XRD patterns of cubic Zn$_3$N$_2$ and cubic TaN according to literature.[40,41] An additional peak of low intensity is observed at ~33.4°, which gradually shifts towards higher 2 Theta angles with increasing Zn alloying concentration of Zn$_x$Ta$_{1-x}$N thin films with *x* from 0.3 to 0.75. The observed peak shift might be attributed to the formation of a novel Zn$_x$Ta$_{1-x}$N alloy, but also varying amounts of precipitation which lead to stress in the films. This peak is more pronounced for depositions at a lower nominal substrate temperature (<83 °C). The high vapor pressure of Zn as well as the low enthalpy of formation of TaN facilitate the formation of TaN$_x$ precipitates (see Figure 1). Traces of a TaN secondary phase are observed in Zn$_x$Ta$_{1-x}$N with *x* as high as 0.7 for samples synthesized at 150 °C, as well as up to *x* = 0.6 for samples synthesized at <83 °C. Overall, due to the low crystallinity of the combinatorial Zn$_x$Ta$_{1-x}$N thin films and overlap of potential signals from Zn$_3$N$_2$ as well as TaN precipitates, it is difficult to conclude if a novel ternary nitride phase does indeed form at intermediate alloying concentrations and if so, if it could be isolated. To verify the finding, further characterization is necessary.

**Chemical state analysis**

We recently demonstrated that chemical state analysis based on the Auger parameter concept can provide additional insights when probing phase formation in thin films of poor crystallinity.[37] To augment the phase characterization via X-ray diffraction mapping we employed chemical state analysis using XPS mapping performed on UHV-transferred Zn-Ta-N libraries.

XPS analysis of $Zn_xTa_{1-x}N$ thin films on selected combinatorial library was carried out without breaking the vacuum to minimize surficial oxidation. Special attention was given to the analysis of the AP, which is defined as a sum of core level binding energy ($CL_{BE}$) of a photoelectron and the kinetic energy of the respective Auger electron ($AE_{KE}$). Thus, by definition, the AP is insensitive to errors caused by erroneous charge correction or charging effects, which are commonly observed during the analysis of semiconducting thin films.[39] As a result, the AP parameter is extremely sensitive to changes in the local chemical environment, including changes in coordination as well as covalency of the bonds.[42] This unique feature makes the AP particularly useful for probing even minute changes in the short-range chemical environment of a given atom.[37,42] To perform a complete chemical state analysis (*i.e.* including cations and anions) we recorded Zn 2p, Ta 4f, N1s core levels as well as Zn LMM, Ta NNN, N KLL Auger spectra for calculating the APs of all constituent elements. The results are summarized in the form of Wagner plots in Fig. 3a-c. The AP for relevant reference samples is given for comparison. The evolutions of the APs of Zn, Ta, as well as N, are plotted as a function of the $Zn_xTa_{1-x}N$ alloying concentration in Fig. 3d-f. This representation is best to highlight changes in the binding environment of the atoms as a function of composition.

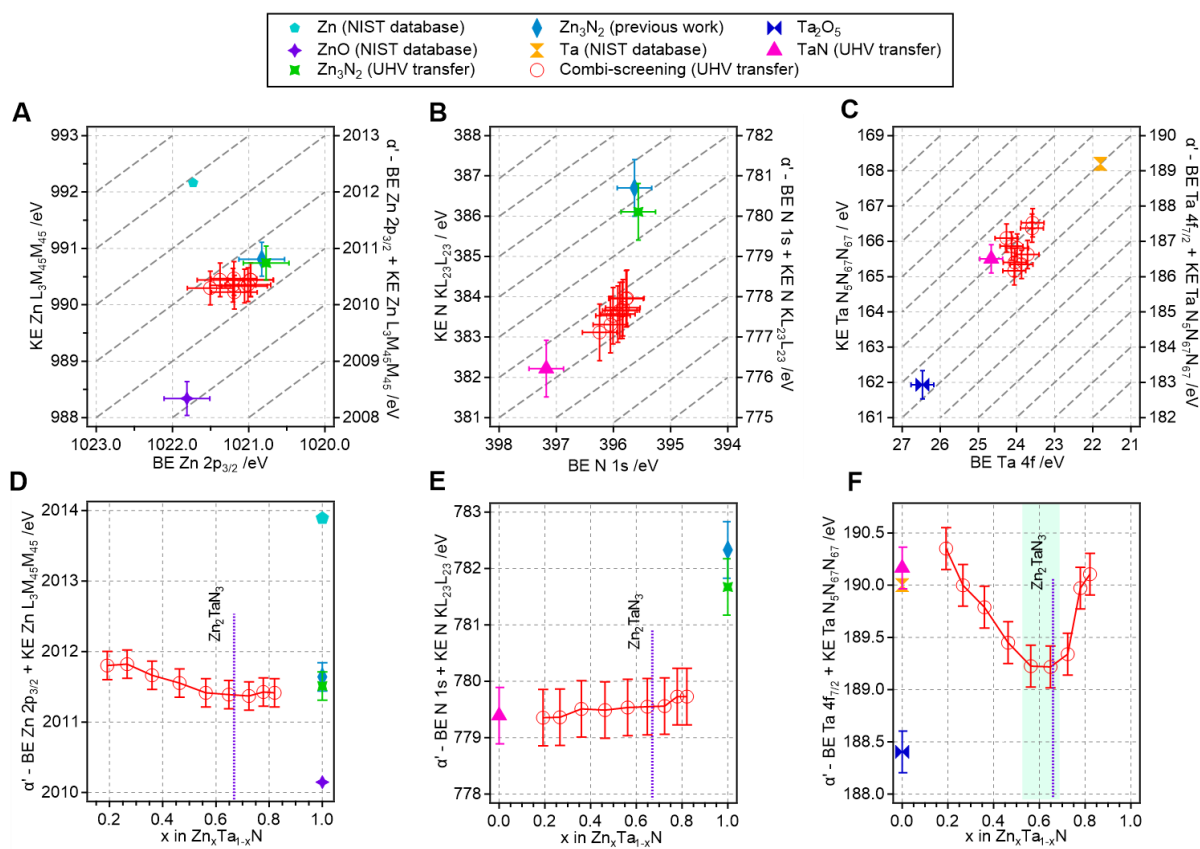

**Figure 3:** XPS study of $Zn_xTa_{1-x}N$ thin films transferred from the sputtering chamber without breaking a high vacuum. Wagner plots for AP based on (a) Zn 2p, (b) N 1s, and (c) Ta 4f core levels. AP of (d) Zn, (e) N, (f) Ta plotted versus composition of $Zn_xTa_{1-x}N$ thin films. Reference measurements on $Zn_3N_2$, TaN, and $Ta_2O_5$ were conducted as part of this work.

Fig. 3a shows the Wagner plot for Zn. In addition to the measurements on the combinatorial libraries, reference measurements on $Zn_3N_2$, as well as TaN thin films, transferred in high vacuum, were conducted since no literature values of sufficient quality were available. XRD patterns of the reference samples are provided in the supporting information (Fig S2). The results for $Zn_3N_2$ are in good agreement with the Zn AP value of $Zn_3N_2$ transported in inert gas atmosphere reported in our previous study,[37] whereas the slightly lower value measured for $Zn_3N_2$ sample transferred in high vacuum is explained by a lower surface oxidation. The Zn 2p core level shows only minor shifts, but considering the Zn AP, the nitride, oxide, and metallic phases are clearly distinguishable. Fig. 3d shows the Zn AP evolution as a function of alloying. Whereas a minimum value is reached for a composition range corresponding to $0.65 \leq x \leq 0.72$ (see Table S1), overall the changes in the Zn AP are minute. This is expected, since in both considered nitride phases (*i.e.* $Zn_3N_2$ and $Zn_2TaN_3$) Zn exhibits a 2+ oxidation state as

S. ZHUK, et al. 2023    Final Version at: doi.org/10.1021/acs.chemmater.3c01341    11

well as tetrahedral coordination (*i.e.* in both structures Zn is surrounded by four nitrogen atoms).[37] Similar to the evolution of the Zn AP, the change in N AP value is also small (Fig. 3b). No apparent change can be observed when the evolution of N AP is plotted as a function of composition, as shown in Fig. 3e. The pronounced jump in AP when moving from Zn$_x$Ta$_{1-x}$N to Zn$_2$N$_3$ can be explained by a change in the overall peak shape in the N KLL emission (see Fig S1). This results in a different feature being attributed as the maximum position. Resolving and fitting the many individual components of the N KLL auger emission could potentially remedy this issue, but it is out of the scope of this study.

The most notable changes are certainly obtained for Ta. As shown in Fig. 3c, in contrast to Zn 2p core level spectra, Ta 4f core level peaks demonstrate a more distinct shift in BE scale. Most strikingly, Fig. 3f shows a high sensitivity of Ta AP to changes in the composition of Zn$_x$Ta$_{1-x}$N thin films. In contrast to the Zn AP evolution, a pronounced minimum can be observed in the compositional region where the novel Zn$_2$TaN$_3$ phase is expected to form (Fig. 3f). Outside of this minimum the Ta AP closely matches the value of δ-TaN. This can be explained by a distinct change in the local chemical environment of the Ta atom in both δ-TaN and the predicted wurtzite-derived Zn$_2$TaN$_3$ phase. In δ-TaN, the Ta atoms exhibit a 3+ oxidation state and 6-fold coordination in edge-sharing octahedra. In wurtzite, Zn$_2$TaN$_3$ Ta exhibits a 5+ oxidation state and tetrahedral coordination. This distinct change in the binding environment can easily explain the pronounced shift in the AP of over 1 eV. These results indicate that a new phase does form in a narrow stability region. Outside of this region, TaN precipitation seems to occur, which is in good agreement with the initial XRD screening.

**Table 1** summarizes important findings from the Auger parameter study on Zn$_2$TaN$_3$, as well as the reference measurements on Zn$_2$N$_3$, and TaN. To the best of our knowledge, these are the first reports of a full chemical state analysis (*i.e.* AP for all constituent elements) for these materials without ambient exposure.

**Table 1:** Core level binding energies (CL$_{BE}$) of photoelectrons, kinetic energies of Auger electrons (AE$_{KE}$), and Auger parameter values for Zn$_3$N$_2$, TaN, and Zn$_{0.65}$Ta$_{0.35}$N.

| | Zn 2p$_{3/2}$ CL$_{BE}$ /eV | Zn (2p$_{3/2}$–2p$_{1/2}$) /eV | Ta 4f$_{7/2}$ CL$_{BE}$ /eV | Ta (4f$_{7/2}$–4f$_{5/2}$) /eV | N 1s CL$_{BE}$ /eV | Zn L$_3$M$_{45}$M$_{45}$ AE$_{KE}$ /eV | Ta N$_5$N$_{67}$N$_7$ AE$_{KE}$ /eV | N KL$_{23}$L$_{23}$ AE$_{KE}$ /eV | Zn AP /eV | Ta AP /eV | N AP /eV |
|---|---|---|---|---|---|---|---|---|---|---|---|
| Zn$_{0.65}$Ta$_{0.35}$N | 1021.1 ±0.1 | 23.06 | 23.9 ±0.1 | 1.88 | 395.9 ±0.1 | 990.3 ±0.1 | 165.4 ± 0.2 | 383.7 ±0.5 | 2011.4 ±0.2 | 189.3 ±0.2 | 779.6 ±0.5 |
| Zn$_3$N$_2$ | 1020.8 ±0.1 | 23.03 | | | 395.6 ±0.1 | 990.7 ±0.1 | | 386.1 ±0.5 | 2011.5 ±0.2 | | 781.7 ±0.5 |
| TaN | | | 24.7 ±0.1 | 1.76 | 397.2 ±0.1 | | 165.5 ± 0.2 | 382.2 ±0.5 | | 190.2 ±0.2 | 779.4 ±0.5 |

Fig. S1 presents the detailed XPS characterization of *in-situ* Zn$_{0.65}$Ta$_{0.35}$N thin films. Despite the high-vacuum transfer, a weak C 1s signal is present, allowing us to align the binding energy scale using the main component of adventitious carbon. Note that, while freshly transferred samples are free of surface oxides, a minor oxidation of the surface was observed during prolonged XPS measurements inside the XPS measurement chamber. Even after longer measurements, the O 1s intensity remains low, and the total amount of oxygen in the surface region does not exceed 10 at. %. Similarly, no oxide components are observed in the Zn 2p and Ta 4f core level spectra, which is important for a meaningful chemical state analysis of nitrides. The VB spectrum of Zn$_2$TaN$_3$ shows a valence band maximum at around 1.1 eV w.r.t the Fermi level, indicating no significant doping of the material.

**Isolation of single-phase Zn$_2$TaN$_3$**

Following the initial screening and discovery of the novel Zn$_2$TaN$_3$ phase via chemical state analysis, more efforts were made to isolate the new phase to improve its phase purity and crystallinity. To this end, additional depositions were carried out using substrate rotating while optimizing the deposition conditions. Fig. 4 shows the XRD characterization of a single composition sample with close to stoichiometric Zn$_2$TaN$_3$ composition. For comparison, the powder patterns for the predicted orthorhombic Zn$_2$TaN$_3$ (SG 36) as well as an experimentally refined powder pattern for wurtzite Zn$_2$TaN$_3$ (SG 186) are displayed in addition to the binary endmember structures.

The Bragg-Brentano XRD measurements show only the (002) and (101) reflections of the WZ structure, indicating a bi-fiber out-of-plane texture.[43] The strong bi-fiber out-of-plane texture is confirmed by pole figure analysis on the (002) as well as (101) reflections (see Figure 4b,c). Both pole figures show a pronounced peak along the substrate normal along with a closed ring at approximately χ = 60°, indicating random in-plane orientation. These results are further supported by grazing incidence wide angle X-ray scattering (GI-WAXS) measurements performed at the European Synchrotron Radiation Facility (ESRF) synchrotron (see Figure S2). To collect the reflections outside of the diffraction-plane, additional measurements were performed while changing χ in 10° increments. The resulting patterns were combined after correcting for the geometry-specific intensity variations. The resulting pattern reveals the remaining reflections associated with the WZ reference powder pattern. It is apparent that the wurtzite reference pattern (SG186) provides a better fit to the experimentally determined XRD patterns compared to the predicted orthorhombic structure (SG36), in particular since no peaks at low diffraction angles are recorded. This is in line with reports in literature. Specifically, several Zn-M-N nitrides have been reported to crystallize in a wurtzite-derived structure, which is closely related to the predicted orthorhombic structure. This results directly from a pronounced cation disorder in the system.[6,18,21]

Using the experimental XRD measurements on $Zn_2TaN_3$ we fitted a unit cell in wurtzite structure (SG 186) based on a previously published structure of the closely related material $ZnGeN_2$ assuming a completely random cation distribution with a Zn/Ta occupancy ratio of 2/1 for each cation site.[44] The lattice parameters of the novel wurtzite $Zn_2TaN_3$ phase are estimated to be approximately a = 3.395 Å and c = 5.40 Å corresponding to a close to ideal c/a ratio of 1.59. The fits were performed using MAUD software,[45] details are given in the supporting information (Figure S3). A detailed refinement of the atomic positions was not possible due to the strong texture of the film.

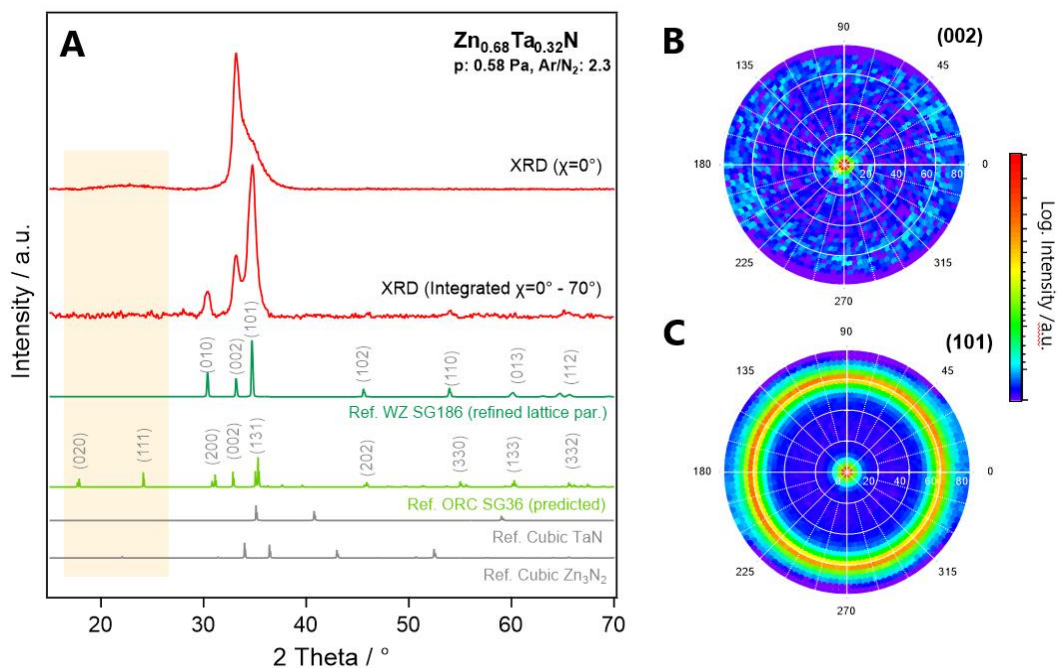

**Figure 4:** a) XRD characterization on a single-composition sample with a nominal composition of $Zn_{0.68}Ta_{0.32}N$. XRD reveals a pronounced out-of-plane texture. Integration over a large χ-range shows all relevant reflections of WZ $Zn_2TaN_3$. b, c) Pole-figure analyses along the (002) and (101) indicate a biaxial out-of-plane texture with random in-plane orientation.

To further improve the crystalline quality of the films deposition on different substrates was attempted, namely c- and r-cut $Al_2O_3$ as well as single crystals of ZnO and MgO. However, no improvement in the crystalline quality was observed.

To analyze the microstructure of the $Zn_2TaN_3$ thin films cross-sectional TEM analysis was performed on a $Zn_xTa_{1-x}N$ sample with a nominal composition of $Zn_{0.68}Ta_{0.32}N$. The results of the analysis are displayed in **Figure 5**. STEM-HAADF as well as bright-field TEM reveal a compact microstructure with vertically oriented grains throughout the specimen. The elemental distribution was studied using STEM-EDS. Elemental Zn, Ta, and N maps show no apparent intensity fluctuations indicating a homogenous sample with no visible precipitation of secondary phases. This is in line with the results from SAED. Here all visible features can be assigned to the previously refined structure of WZ $Zn_2TaN_3$. Finally, HR-TEM was performed on an individual grain. The Fast Fourier Transform (FFT) image along the zone axis (110) further confirms the phase assignment and provides a very good fit with the proposed structure (for details see **Figure S5**).

The analyzed sample therefore appears to be phase-pure WZ $Zn_2TaN_3$ (SG 186). Composition analysis of this sample using ERDA (see **Figure S6**) reveals an almost perfectly stoichiometric cation ratio with Zn: 30.1 at. % and Ta: 14.3 at. %, whereas the nitrogen content is slightly higher with N: 54.5 at. %. Furthermore, the content of O in the bulk of the film has been found to be 0.6 at. % although the sample was stored in air for several days prior to the analysis. This confirms the chemical stability of the novel phase in ambient conditions. The low oxygen levels are further confirmed by quantification from STEM EDS (see **Figure S7**). Additional TEM analysis was performed on slightly Ta-rich films with a nominal composition of $Zn_{0.60}Ta_{0.40}N$ (see **Figure S8**). While the results closely match those of the analysis on the phase pure $Zn_{0.68}Ta_{0.32}N$ sample, trace amounts of cubic TaN precipitates can be found as a result of the higher Ta content. This further highlights the very narrow stability window of WZ $Zn_2TaN_3$.

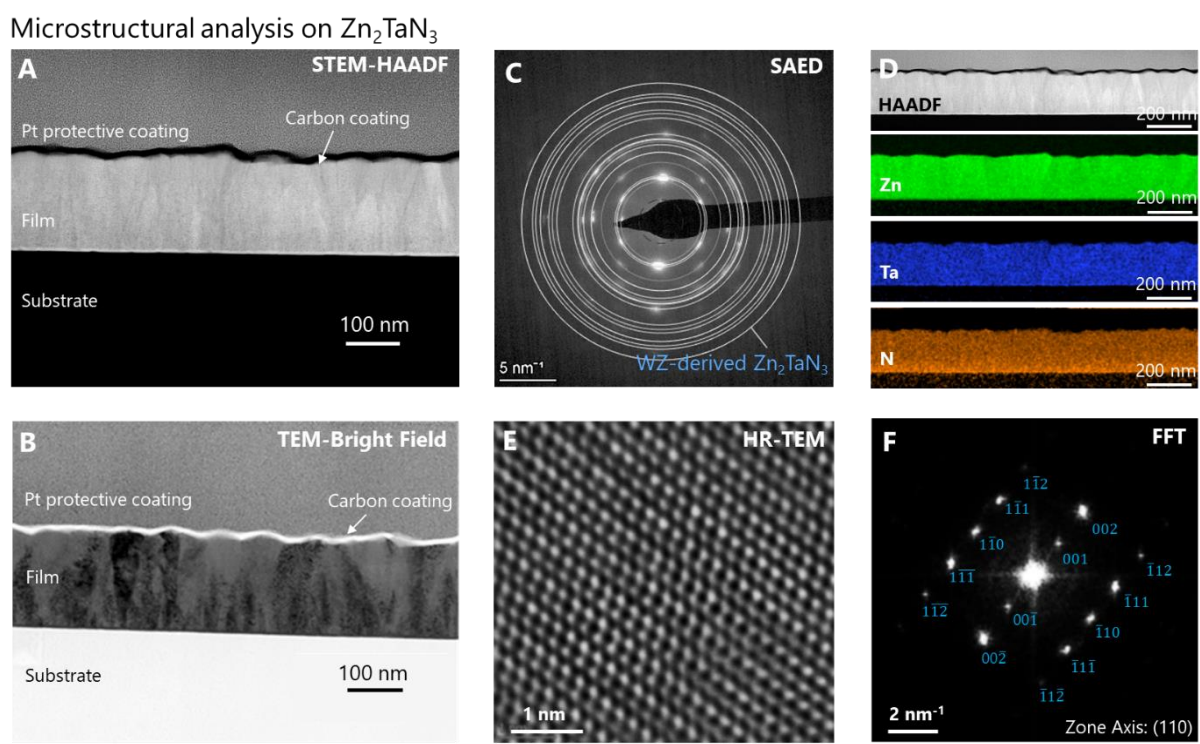

**Figure 5:** Microstructural analysis of a thin film with a nominal composition of $Zn_{0.68}Ta_{0.32}N$. Shown are micrographs using A) STEM-HAADF and B) TEM bright field imaging. C) SAED as well as D) EDS elemental mapping, indicating the formation of wurtzite $Zn_2TaN_3$ with no apparent precipitation of secondary phases. The WZ reference pattern is displayed as theoretical rings in the SAED image. E) HR-TEM as well as the corresponding F) Fourier transformed images further confirm the assigned structure. Details are provided in the supporting information.

S. ZHUK, et al. 2023    Final Version at: doi.org/10.1021/acs.chemmater.3c01341    16

**Mapping of functional properties**

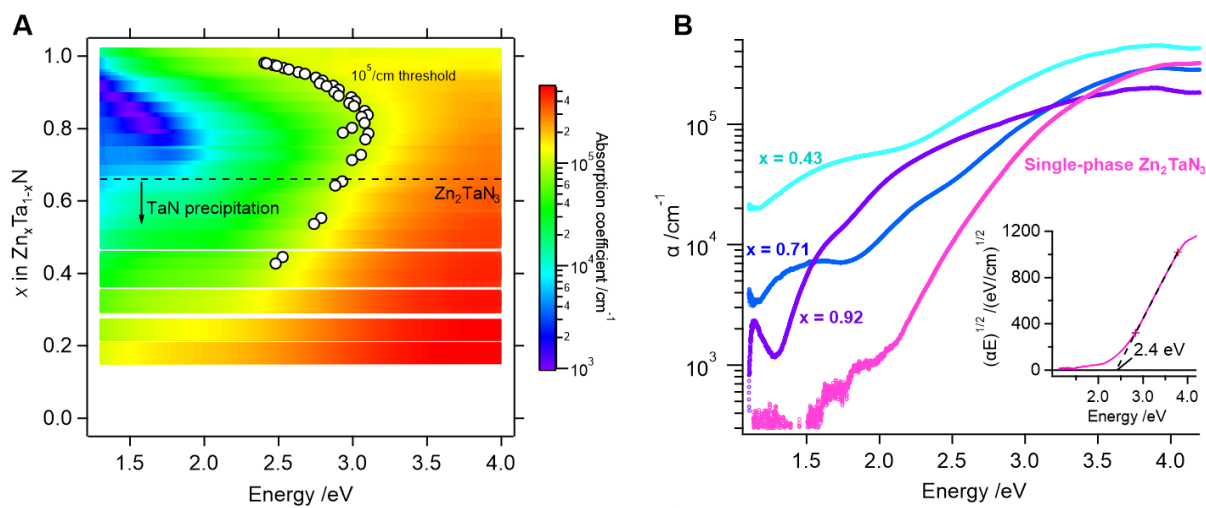

**Figure 6:** a) Absorption coefficient from UV-Vis mapping for $Zn_xTa_{1-x}N$ thin films synthesized at ≤83°C. The $10^5$/cm threshold is displayed as circles. The optical band gap increases for compositions close to $Zn_2TaN_3$, whereas TaN precipitation leads to significant sub-band gap absorption for Ta-rich films. b) Absorption coefficient for selected compositions as well as single-phase $Zn_2TaN_3$. The inset figure shows the Tauc plot for an indirect transition for the single-phase sample.

The optical properties of combinatorial $Zn_xTa_{1-x}N$ thin films were mapped using our in-house built automated high-throughput UV-Vis photo-spectroscopy system. **Figure 6** shows the absorption coefficient of $Zn_xTa_{1-x}N$ thin films plotted as a function of composition and photon energy for samples synthesized at ≤83°C. A significant decrease in absorption can be seen for intermediate alloying concentrations. As a result the $10^5$/cm absorption threshold shifts to higher photon energies. This indicates on one hand a potential change in band gap and on the other hand a potential reduction in absorption and scattering caused by secondary phases. This is particularly noteworthy, as this result, in addition to the chemical state analysis mapping, can also hint at the formation of a different phase at intermediate alloying concentrations. The highest $10^5$/cm absorption onset for $Zn_xTa_{1-x}N$ thin films is found for samples with slightly higher Zn content relative to stoichiometric $Zn_2TaN_3$. This effect is most pronounced for higher synthesis temperatures (see **Figure S9**) and can be explained by sub-bandgap absorption caused by metallic TaN precipitates. This observation is corroborated by the results of XRD screening given in Fig. 2a. Figure 6b shows absorption coefficients for selected concentrations as well as for single-phase $Zn_2TaN_3$. The lack of impurity scattering for the single-phase sample facilitates the measurement of the optical band gap. Here, Tauc analysis indicates an indirect band gap of 2.4 eV (Figure 6b, inset). The optical band gap is notably lower than the prediction of 3.58 eV by Hinuma *et al.*.[19] The reason is likely rooted in the observed cation disorder. A similar reduction in band offset was already observed for $Zn_2VN_3$ as well as $ZnGeN_2$.[14,20] In $ZnGeN_2$, the reduced band gap was reported to be caused by cation anti-site defects.[18,20]

Finally, the electrical properties of $Zn_xTa_{1-x}N$ thin films were characterized using an automated four-point probe measurement system. Figure S10 demonstrates the resistivities of $Zn_xTa_{1-x}N$ thin films synthesized at temperatures of 150°C and ≤83°C. The resistivity of the samples at the ends of the studied composition range is comparatively low, as expected for TaN, and $Zn_3N_2$ phases.[14,46] However, the measured resistivity is comparatively high for $Zn_xTa_{1-x}N$ samples with intermediate alloying concentrations. The values exceed the measurement range of our mapping instrument (i.e. sheet resistances of >20 GΩ☐). Thus, the composition region where wurtzite $Zn_xTa_{1-x}N$ thin films are formed is characterized by a high resistivity of over $10^5$ Ωcm, exceeding the resistivity of TaN and $Zn_3N_2$ phases by at least several orders of magnitude. Less accurate two-point measurements on single-phase $Zn_2TaN_3$ indicate a resistivity well above $10^7$ Ωcm. This is comparable with reports of $ZnGeN_2$, where resistivity values in the range of $10^9$ Ωcm - $10^{10}$ Ωcm have been reported.[47] The high resistivity is in good agreement with the valence band maximum from XPS of around 1.1 eV with respect to the Fermi level which indicates no significant doping. Depending on the desired application alloying on the cation site or controlled off-stoichiometry could potentially be leveraged to increase the doping of the material.

**Conclusion**

An exploration of Zn-Ta-N phase space was carried out using a combinatorial reactive PVD coupled with high-throughput material characterization. Despite ambiguous results from the initial XRD phase screening, chemical state analysis mapping indicated a clear discontinuity in the local Ta chemical environment, in line with the formation of the new phase. Further experiments verified the formation of the new phase and additional synthesis efforts resulted in the first reported synthesis of single-phase $Zn_2TaN_3$ in wurtzite structure, confirming earlier predictions by Zakutayev *et al.* [6] and Hinuma *et al.*.[19] Combined STEM and XRD characterization of $Zn_2TaN_3$ thin films confirmed the phase-purity and structure of the thin films. The functional properties indicate that $Zn_2TaN_3$ is a wide band gap semiconductor with an indirect band gap of 2.4 eV and high resistivity of $>10^5$ Ω cm. The practicality for optoelectronic applications however might be challenging due to the identified narrow stability window of the novel phase. Finally, as part of this study, we were able to present previously unreported AP measurements for several materials including $Zn_3N_2$ and TaN.

Overall, the results presented in this work highlight common challenges in the screening of new materials and demonstrate the value of adding materials characterization techniques sensitive to changes in the local binding environment. Complementing combinatorial workflows with such techniques will enable researchers to screen new phase spaces with higher confidence and apply more targeted efforts toward the development of new functional materials that might otherwise be overlooked.

## Supporting Information

The supporting information contains additional experimental data: Detailed XPS data (Table S1); Detailed XP spectra (Figure S1); XRD measurements of TaN and $Zn_3N_2$ reference samples (Figure S2); Details regarding the refinement of the lattice parameters (Figure S3); GI-WAXS measurements on single-phase $Zn_2TaN_3$ (Figure S4); Details regarding the SAED and FFT indexing (Figure S5); ERDA analysis (Figure S6); STEM EDS analysis of the O contamination in the films (Figure S7); TEM analysis of $Zn_{0.6}Ta_{0.4}N$ (Figure S8); UV-Vis mapping for different deposition temperatures (Figure S9); Resistivity mapping data (Figure S10).

## Acknowledgments

S.Z. acknowledges research funding from Empa Internal Research Call 2020. J. P. acknowledges funding by the SNSF (project no. 200021_196980). A.W. acknowledges funding from the Strategic Focus Area–Advanced Manufacturing (SFA–AM) through the project Advancing manufacturability of hybrid organic–inorganic semiconductors for large area optoelectronics (AMYS). The authors acknowledge Simon Christian Böhme and Maksym Kovalenko from the Laboratory of Inorganic Chemistry at ETH Zürich for their help with cryo photoluminesce analysis (not reported). A.W. further acknowledges Dmitry Chernyshov for the introduction and help at the Swiss-Norwegian BeamLines located at the ESRF as well as Christian Wolff from EPF Lausanne for sharing the measurement time. The authors thank Arnold Müller and Christof Vockenhuber from the Laboratory of Ion Beam Physics at ETH Zurich for RBS and ERDA analysis.

**Author contributions (CRediT):**

Siarhei Zhuk: Investigation, Formal Analysis, Validation, Writing-Original Draft, Visualization, Writing – Review & Editing; Alexander Wieczorek: Investigation, Software, Formal Analysis, Writing – Review & Editing; Amit Sharma: Investigation, Formal Analysis, Writing – Review & Editing; Jyotish Patidar: Investigation, Writing – Review & Editing; Kerstin Thorwarth: Investigation, Writing – Review & Editing; Johann Michler: Supervision, Writing – Review & Editing; Sebastian Siol: Investigation, Formal Analysis, Writing-Original Draft, Visualization, Supervision, Project Administration, Funding Acquisition, Writing – Review & Editing

# Supplementary Information:

# Combinatorial Reactive Sputtering with Auger Parameter Analysis Enables Synthesis of Wurtzite Zn$_2$TaN$_3$.

Siarhei Zhuk[a], Alexander Wieczorek[a], Amit Sharma[b], Jyotish Patidar,[a]

Kerstin Thorwarth[a], Johann Michler[b], Sebastian Siol[a,*]

a) Empa – Swiss Federal Laboratories for Materials Science and Technology,
8600 Dübendorf, Switzerland

b) Empa – Swiss Federal Laboratories for Materials Science and Technology,
3602 Thun, Switzerland

*E-mail: sebastian.siol@empa.ch

**Table S1** – Raw data of the core-level binding energies of photoelectrons, kinetic energies of Auger electrons and Auger parameter values for in-situ $Zn_3N_2$, TaN, and $Zn_xTa_{1-x}N$ thin films. The typical inaccuracy of the measured values is 0.2 eV due to the C1s charge correction. The typical error in determining the BE for core level spectra is 0.1 eV, for the N KLL in particular the broad peak leads to a larger error of approximately 0.5 eV.

| | Zn $2p_{3/2}$ $CL_{BE}$ /eV | Zn ($2p_{3/2}$–$2p_{1/2}$) /eV | Ta $4f_{7/2}$ $CL_{BE}$ /eV | Ta ($4f_{7/2}$–$4f_{5/2}$) /eV | N 1s $CL_{BE}$ /eV | Zn $L_3M_{45}M_{45}$ $AE_{KE}$ /eV | Ta $N_5N_{67}N_7$ $AE_{KE}$ /eV | N $KL_{23}L_{23}$ $AE_{KE}$ /eV | Zn AP /eV | Ta AP /eV | N AP /eV |
|---|---|---|---|---|---|---|---|---|---|---|---|
| $Zn_{0.19}Ta_{0.81}N$ | 1021.5 | 23.04 | 24.26 | 1.88 | 396.24 | 990.3 | 166.09 | 383.11 | 2011.8 | 190.35 | 779.35 |
| $Zn_{0.27}Ta_{0.73}N$ | 1021.38 | 23.04 | 24.13 | 1.9 | 396.05 | 990.44 | 165.87 | 383.31 | 2011.82 | 190 | 779.36 |
| $Zn_{0.36}Ta_{0.64}N$ | 1021.2 | 23.06 | 23.98 | 1.88 | 395.85 | 990.47 | 165.81 | 383.67 | 2011.67 | 189.79 | 779.52 |
| $Zn_{0.46}Ta_{0.54}N$ | 1021.21 | 23.08 | 24.05 | 1.92 | 395.92 | 990.35 | 165.4 | 383.57 | 2011.56 | 189.45 | 779.49 |
| $Zn_{0.56}Ta_{0.44}N$ | 1021.19 | 23.08 | 24.06 | 1.87 | 396.01 | 990.22 | 165.17 | 383.53 | 2011.41 | 189.23 | 779.54 |
| $Zn_{0.65}Ta_{0.35}N$ | 1021.06 | 23.06 | 23.87 | 1.88 | 395.89 | 990.33 | 165.35 | 383.66 | 2011.39 | 189.22 | 779.55 |
| $Zn_{0.72}Ta_{0.28}N$ | 1021.01 | 23.09 | 23.71 | 1.87 | 395.84 | 990.36 | 165.63 | 383.73 | 2011.37 | 189.34 | 779.57 |
| $Zn_{0.78}Ta_{0.22}N$ | 1020.98 | 23.06 | 23.6 | 1.89 | 395.77 | 990.45 | 166.38 | 383.96 | 2011.43 | 189.98 | 779.73 |
| $Zn_{0.82}Ta_{0.18}N$ | 1020.97 | 23.08 | 23.58 | 1.88 | 395.78 | 990.44 | 166.53 | 383.95 | 2011.41 | 190.11 | 779.73 |
| $Zn_3N_2$ | 1020.77 | 23.03 | | | 395.57 | 990.74 | | 386.11 | 2011.51 | | 781.68 |
| TaN | | | 24.66 | 1.76 | 397.17 | | 165.5 | 382.22 | | 190.16 | 779.39 |

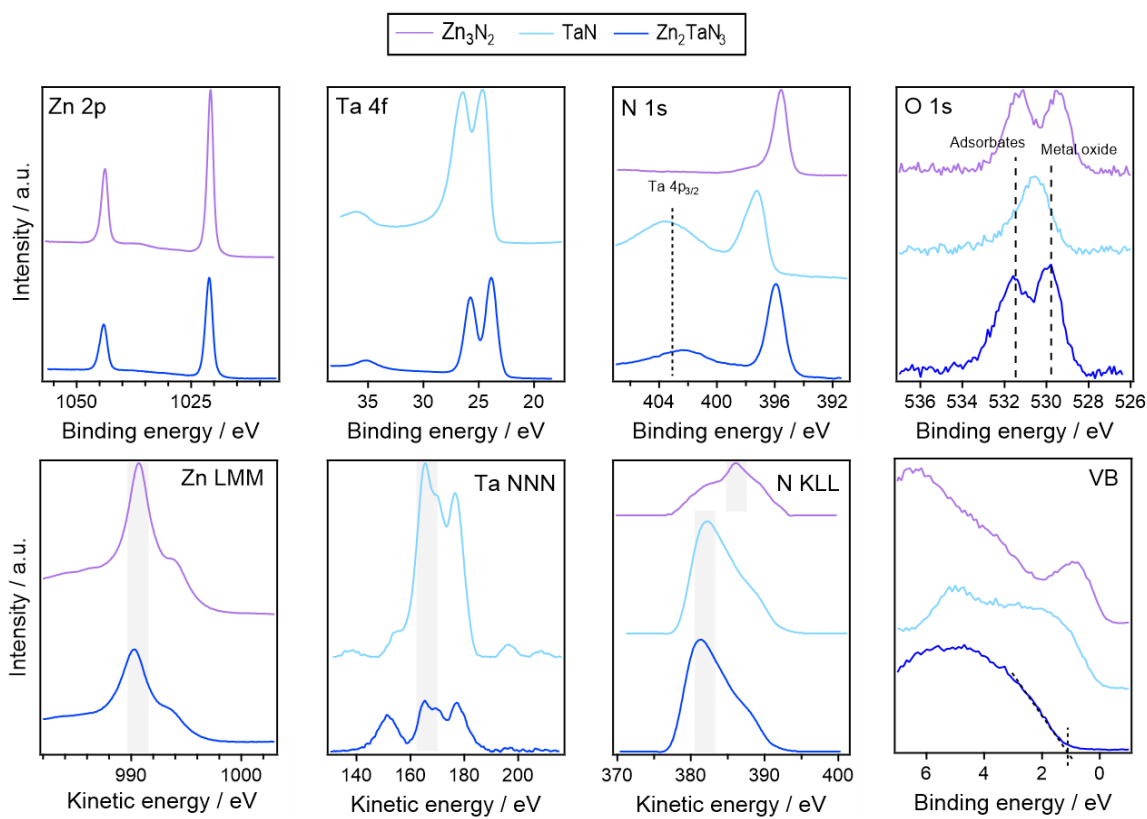

**Fig. S1.** XPS characterization of Zn 2p, Ta 4f, N1s, and O 1s core level spectra as well as the corresponding Zn LMM, Ta NNN, N KLL Auger lines, and the valence band (VB) emission for the binaries $Zn_3N_2$, and TaN as well as stoichiometric $Zn_2TaN_3$. The VBM of $Zn_2TaN_3$ was determined to be at approximately 1.1 eV w.r.t. the Fermi level indicating no significant intrinsic doping. The samples were transferred in UHV conditions. During longer measurements, residual oxygen in the XPS measurement chamber leads to minimal surface contamination. For all measurements, the surface oxygen content was less than 10 at.%.

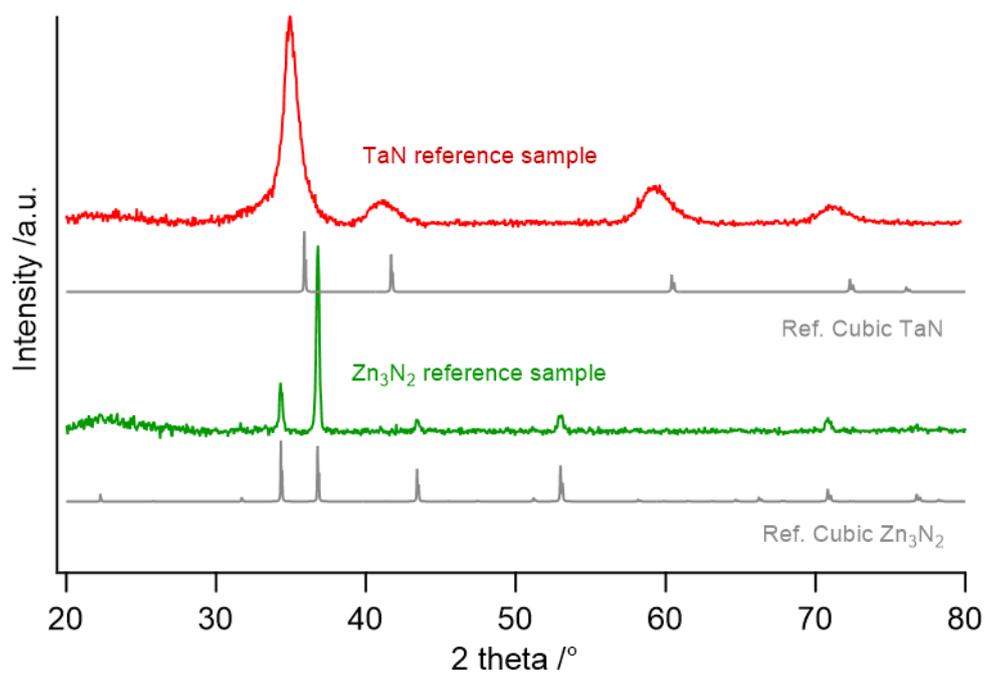

**Fig. S2.** XRD analysis on samples produced with comparable deposition conditions as the films used to produce the reference measurements for the XPS study. The XRD patterns confirm the targeted cubic TaN and $Zn_3N_2$ phases, respectively.

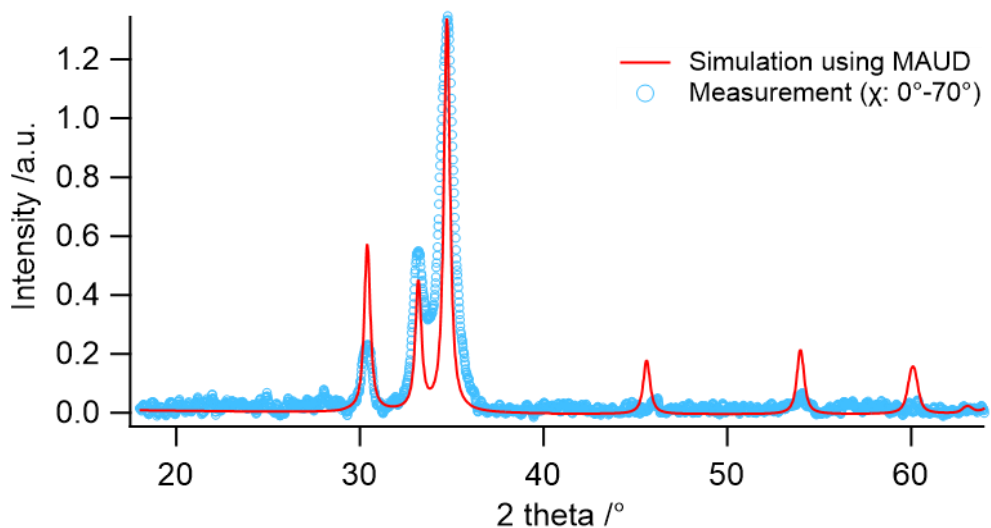

**Fig. S3.** Simulated XRD pattern used for the lattice parameter refinement. The underlying structure was ZnGeN$_2$ in wurtzite structure. The lattice parameters of wurtzite Zn$_2$TaN$_3$ were estimated to be approximately a = (3.395 ±0.01) Å and c = (5.40 ±0.01) Å corresponding to a close to ideal c/a ratio of 1.59. The fitting was performed using MAUD software after determining the instrument misalignment and peak broadening using a corundum reference sample. For the simulation an average grain size of 30 nm was assumed. Due to the strong texture and low grain size of the sample no refinement of the atomic positions was performed.

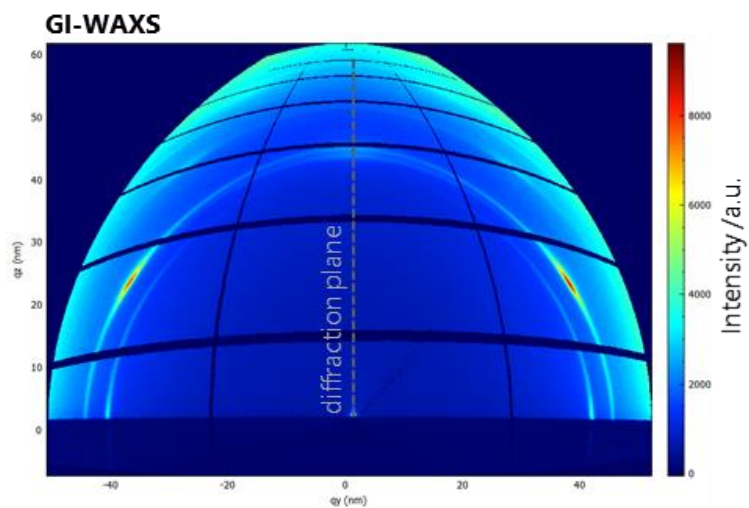

**Fig. S4.** GI-WAXS measurements collected on the single-phase $Zn_2TaN_3$ sample. The image clearly shows the preferential out-of-plane orientation of the grains and explains the low intensity when measured in Bragg-Brentano XRD (*i.e.* when only reflections in the diffraction plane are collected). Synchrotron X-ray diffraction was performed at the BM01 diffraction beamline of the Swiss-Norwegian BeamLines (SNBL) of the European Synchrotron Radiation Facility (ESRF) in Grenoble, France. The wavelength during measurements was 0.694 ˚A. A Pilatus 2M2D detector was used. For details, see: Dyadkin *et al.*, "A new multipurpose diffractometer PILATUS@SNBL". Journal of Synchrotron Radiation 2016, 23, 825–829

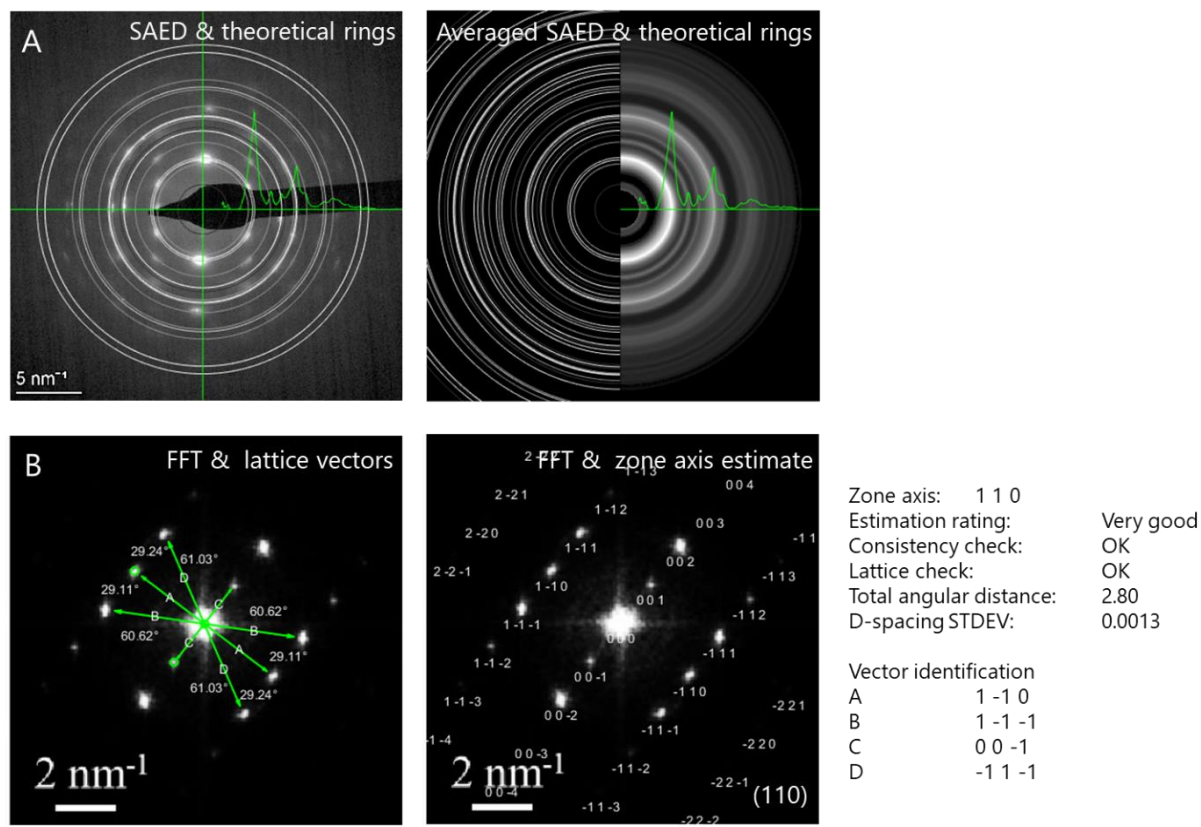

**Fig. S5.** Detailed report of the indexing performed on the SAED (a) and FFT (b) images recorded on the single-phase $Zn_2TaN_3$ sample. All recorded reflections can be assigned to a wurtzite unit cell (SG186) with lattice parameters a = 3.395 Å and c = 5.40 Å. An assignment to the orthorhombic structure (SG36) was also possible, but lead to a higher mismatch. This further highlights how closely related the two structures are. The analysis was performed using the software CrysTBox ringGUI 1.16 as well as CrysTBox diffractGUI 2.21 by Miloslav Klinger.

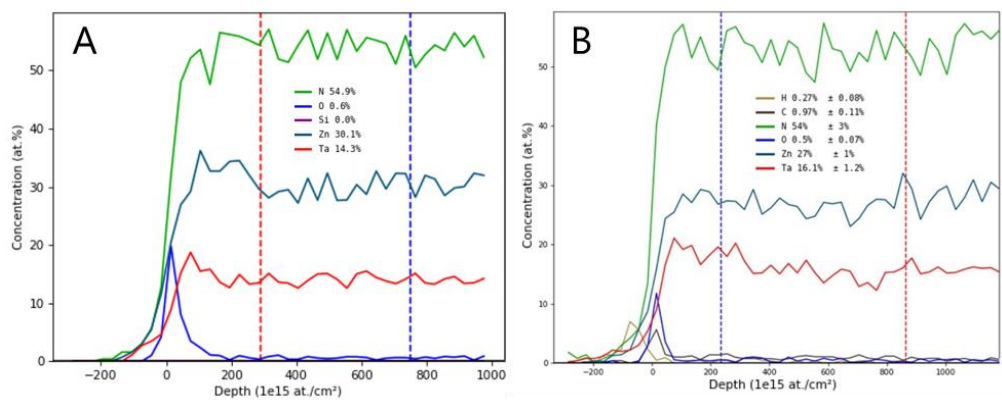

**Fig. S6.** ERDA depth profiles of the two $Zn_xTa_{1-x}N$ thin films on EXG used for the detailed TEM analysis. The bulk oxygen content is well below 1 at.% for both samples: a) single-phase $Zn_2TaN_3$, b) $Zn_{0.6}Ta_{0.4}N$.

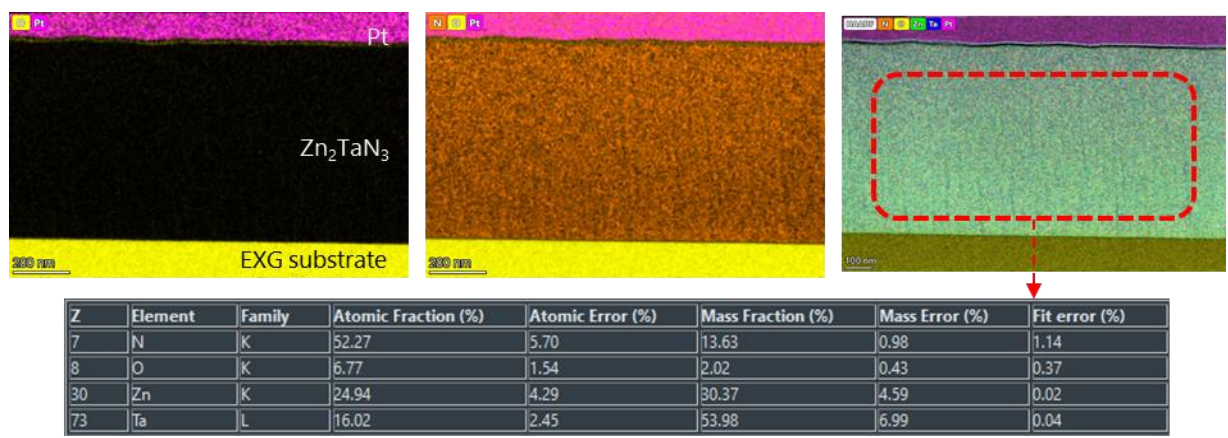

**Fig. S7.** STEM EDS analysis of the oxygen content of a sample with a nominal composition of $Zn_{0.6}Ta_{0.4}N$. The FIB lamella was handled in air leading to surface contamination/oxidation on both sides. Still, the oxygen contamination is < 7 at.% confirming the compositional analysis from ERDA.

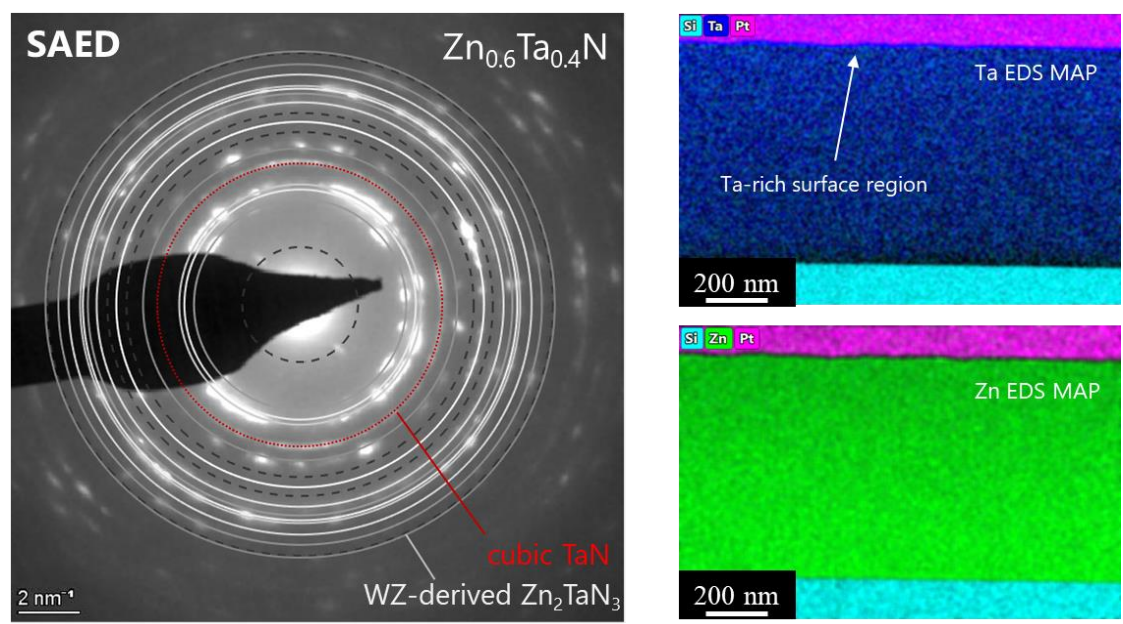

**Fig. S8.** TEM analysis on a sample with a nominal composition of $Zn_{0.6}Ta_{0.4}N$ (Fig. S3b). The higher Ta content leads to precipitation of cubic TaN as evidenced by SAED and EDS elemental mapping.

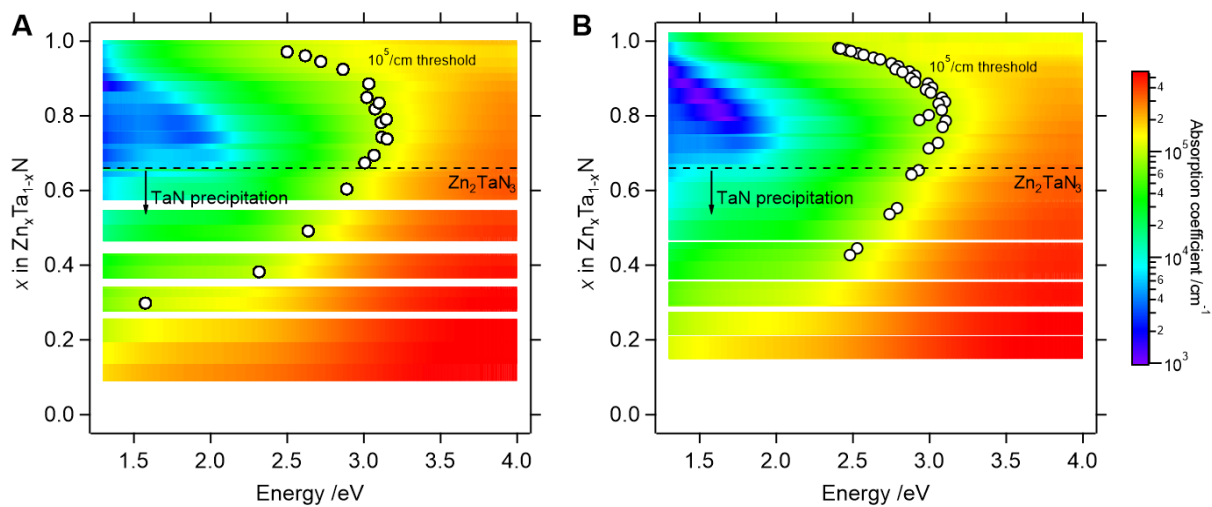

**Figure S9:** Absorption coefficient of Zn$_x$Ta$_{1-x}$N thin films synthesized at temperatures of 150°C and ≤83°C. TaN precipitates lead to significant sub-bandgap absorption for Ta-rich films.

**Details regarding the optical characterization**

Absorption coefficient (α) was calculated as α = 2.033 $A/d$ where $d$ is film thickness and $A$ is absorbance calculated using the following formula:

$$A = -\log(T_{Substrate+Film}/(1-R_{Substrate+Film})/(T_{Substrate}/(1-R_{Substrate})))$$

There $R$ and $T$ denote reflectance and transmittance, respectively.

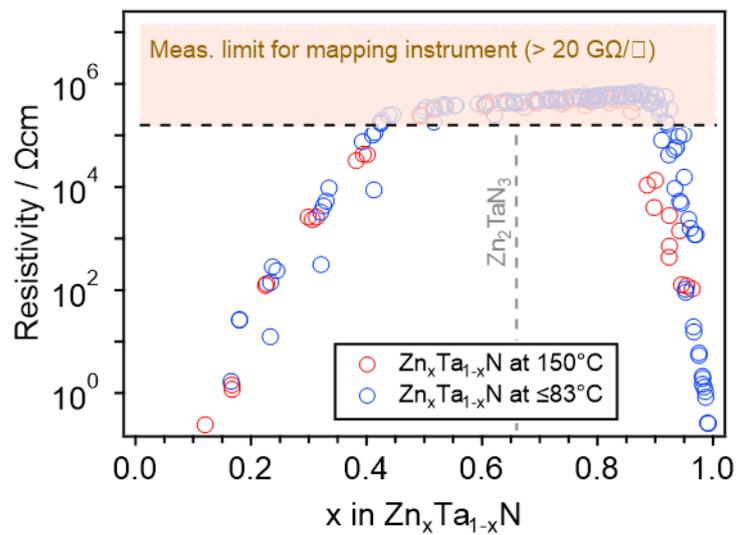

**Figure S10:** Resistivity mapping of $Zn_xTa_{1-x}N$ thin films synthesized at temperatures of 150°C and ≤83°C.